# Local Terahertz Field Enhancement for Time-Resolved X-ray Diffraction

M. Kozina,[1] M. Pancaldi[2,3], C. Bernhard,[4] T. van Driel,[1] J.M. Glownia,[1] P. Marsik,[4] M. Radovic,[5] C.A.F. Vaz,[5] D. Zhu,[1] S. Bonetti,[3] U. Staub,[5] and M.C. Hoffmann[1]

[1]*Linac Coherent Light Source, SLAC National Accelerator Laboratory, Menlo Park, California 94025, USA*

[2]*CIC nanoGUNE, E-20018 Donostia-San Sebastian, Spain*

[3]*Department of Physics, Stockholm University, SE-106 91 Stockholm, Sweden*

[4]*Department of Physics, University of Fribourg, Chemin du Musée 3, CH-1700 Fribourg, Switzerland*

[5]*Swiss Light Source, Paul Scherrer Institut, 5232 Villigen PSI, Switzerland*

**Abstract:** We report local field strength enhancement of single-cycle terahertz (THz) pulses in an ultrafast time-resolved x-ray diffraction experiment. We show that patterning the sample with gold microstructures increases the THz field without changing the THz pulse shape or drastically affecting the quality of the x-ray diffraction pattern. We find a five-fold increase in THz-induced x-ray diffraction intensity change in the presence of microstructures on a $SrTiO_3$ thin-film sample.

Femtosecond x-ray sources provide unique insight into the dynamics of matter on ultrafast timescales. In particular, the combination of the high brightness and sub-100 femtosecond pulse duration of the Linac Coherent Light Source (LCLS)[1] with various ultrafast radiation sources (e.g. visible, ultraviolet, or terahertz) employed in a pump-probe experiment has enabled exploration of dynamics in atoms and molecules[2–4], semiconductors[5,6], and correlated electron systems [7–11]. The majority of pump-probe experiments performed thus far have relied on intensity driven processes[8,12,13] where dynamics are induced by the impulsive nature of the pump excitation. In contrast, excitation with terahertz (THz) pulses provides access to new dynamics driven by the field of the pump radiation[14–17]. Single-cycle THz pulses are intrinsically carrier-envelope-phase stable, and so repeated excitation with this radiation in a pump-probe experiment maintains not only the temporal intensity profile but also the phase of the electric field, thereby enabling a true field-driven excitation. Moreover, the pulse duration of the THz is sufficiently long compared to the probing x rays (1 ps compared to 30 fs), allowing x-ray interrogation of the THz-induced dynamics without loss of temporal resolution.

Current single-cycle THz radiation sources based on Ti:Sapphire ultrafast lasers are limited to several MV/cm peak electric field strength[18]. While THz radiation couples strongly to IR active phonon modes in ions with strong dipole moments (e.g. $SrTiO_3$,[19] $BaTiO_3$[17]), in order to explore richer dynamics, for example transient phases of matter or anharmonic coupling, larger THz field strengths are needed. Recently, the development of metamaterials has significantly enhanced the local field of THz radiation[20–25]. In order to make use of this enhanced field in a THz pump, x-ray probe experiment, it is essential that the THz temporal structure remain unchanged and that the metamaterial design not interfere with the x-ray measurement. One solution for metamaterial design satisfying these criteria incorporates patterning a large area of a thin-film sample with metal stripes several micrometers in width and several hundred nanometers thick, and probing with an x-ray spot size that is much larger than the stripe period. Note this approach was first discussed in [26]. The x-ray diffraction from the polycrystalline metal stripes scatters into powder rings that need not overlap in reciprocal space with the sample scattering. Moreover, the thin metal stripes will transmit partly the x rays but block the THz. Therefore it is essential to remove the sample below the metal stripes to ensure that scattering from the sample comes only from regions excited by the THz. We expect the scattering from the film to be reduced by the fraction of sample removed to create the metal stripes. It is helpful to use a thin film rather than a bulk sample because in the thin-film case the x-ray penetration depth is more closely matched to the depth of THz field enhancement provided by the metal stripes. Additionally, the broad diffraction peak from the thin film relaxes normalization constraints and enhances the signal to noise compared to the bulk. Here we describe in detail the enhancement provided by the metamaterial structure to the THz local field and subsequent transient x-ray scattering signal, revealing THz driven ionic motion in SrTiO$_3$.

Our samples consisted of 100 nm thick films of SrTiO$_3$ (STO) grown by pulsed laser deposition on (001) (La$_{0.3}$Sr$_{0.7}$)(Al$_{0.65}$Ta$_{0.35}$)O$_3$ (LSAT) substrate. Under these growth conditions, STO is slightly strained (3.867 Å in-plane matched to bulk LSAT and 3.925 Å out-of-plane, compared to 3.905 Å for bulk STO), leading to a hardening of the soft mode phonon[27]. Thus the bandwidth of the incident THz radiation (DC up to 2.5 THz) was below the lowest zone-center optical phonon frequency at 100 K[27]. We expect even stronger coupling of the THz to the sample when there is significant overlap between the sample optical phonon and THz spectra due to resonant absorption effects.

On one STO film a metal resonator structure was deposited to locally enhance the THz field strength. The resonator structure consisted of a 1.5×1.5 mm$^2$ array of 8.5 μm wide Au stripes with a gap spacing of 1.5 μm, set at 45° from the [100] direction of the STO (001). The Au resonator structures were fabricated using a self-alignment, single step e-beam lithography process. The STO/LSAT (001) was first spin-coated with a ~200 nm methyl methacrylate (MMA) resist layer, a 1.5 nm Cr adhesion layer followed by a ~600 nm hydrogen silsesquioxane (HSQ) resist layer. The negative of the resonator structure was then transferred to the HSQ resist layer using an e-beam writer. Upon development, SiO$_x$ lines rest on the continuous MMA resist layer. The MMA areas not protected by the SiO$_x$ were then removed with O-plasma, leaving a SiO$_x$ mask layer for the subsequent etching of the exposed STO areas using Ar ion beam etching for a total of 140 nm as measured using a profilometer on a shadow-masked area of the sample with no SiO$_x$ mask layer. A 3 nm Cr/115 nm Au film was then deposited by thermal evaporation and a subsequent ultra-sound assisted lift-off in acetone removed the Au/SiO$_x$ parts of the sample. We show in Fig. 1A a cartoon of the sample cross-section and in Fig. 1B an overhead cartoon with the size of the x-ray spot for comparison.

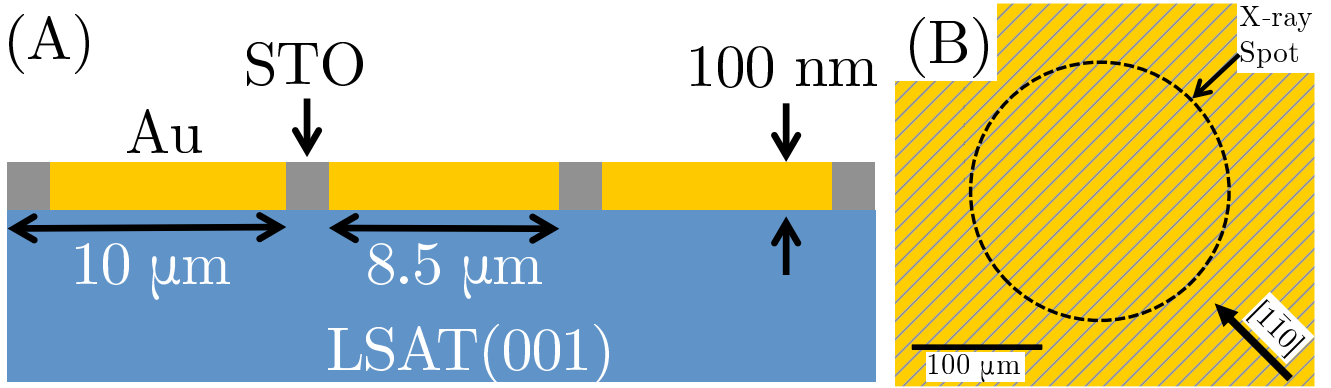

Fig. 1. (A) Schematic cross-section of sample. (B) Schematic surface of sample. The gold stripes are oriented parallel to [110]. The dashed circle represents the x-ray spot on the sample at normal incidence.

We collected time-resolved x-ray diffraction measurements on the STO film under excitation below resonance with single-cycle THz fields at 100 K. By varying the time delay between the THz waveform and the interrogating x-ray pulses, we recorded the transient change in diffraction intensity of the STO film induced by the THz radiation. The THz radiation was generated via optical rectification utilizing the tilted-pulse front technique in LiNbO$_3$[28,29] pumped by an 800 nm laser (20 mJ, 100 fs, 120 Hz). In Fig. 3A, we show electro-optic sampling (EOS) measurements of the THz waveform incident on the sample (black trace) measured with a fraction of the 800 nm light used for generating the THz. The maximum field strength of the incident THz waveform was 900±50 kV/cm (measured via the electro-optic modulation of GaP at the sample position). We performed COMSOL® simulations to calculate the strength of the THz field in the STO with and without the metamaterial structure incorporating the STO dispersion at 100 K obtained from ellipsometry[27] and the measured EOS in free space (Fig. 3A). In Fig. 2A, we show a cross-section simulation of the spatial dependence of the field enhancement across the metal gap at a fixed frequency of 1 THz. Using the measured EOS trace as an input to our simulation, we present the calculated time domain field with (red) and without (black) the metamaterial structure in Fig. 2B. We estimate a peak field enhancement of 5.3 in the presence of the metamaterial antennas.

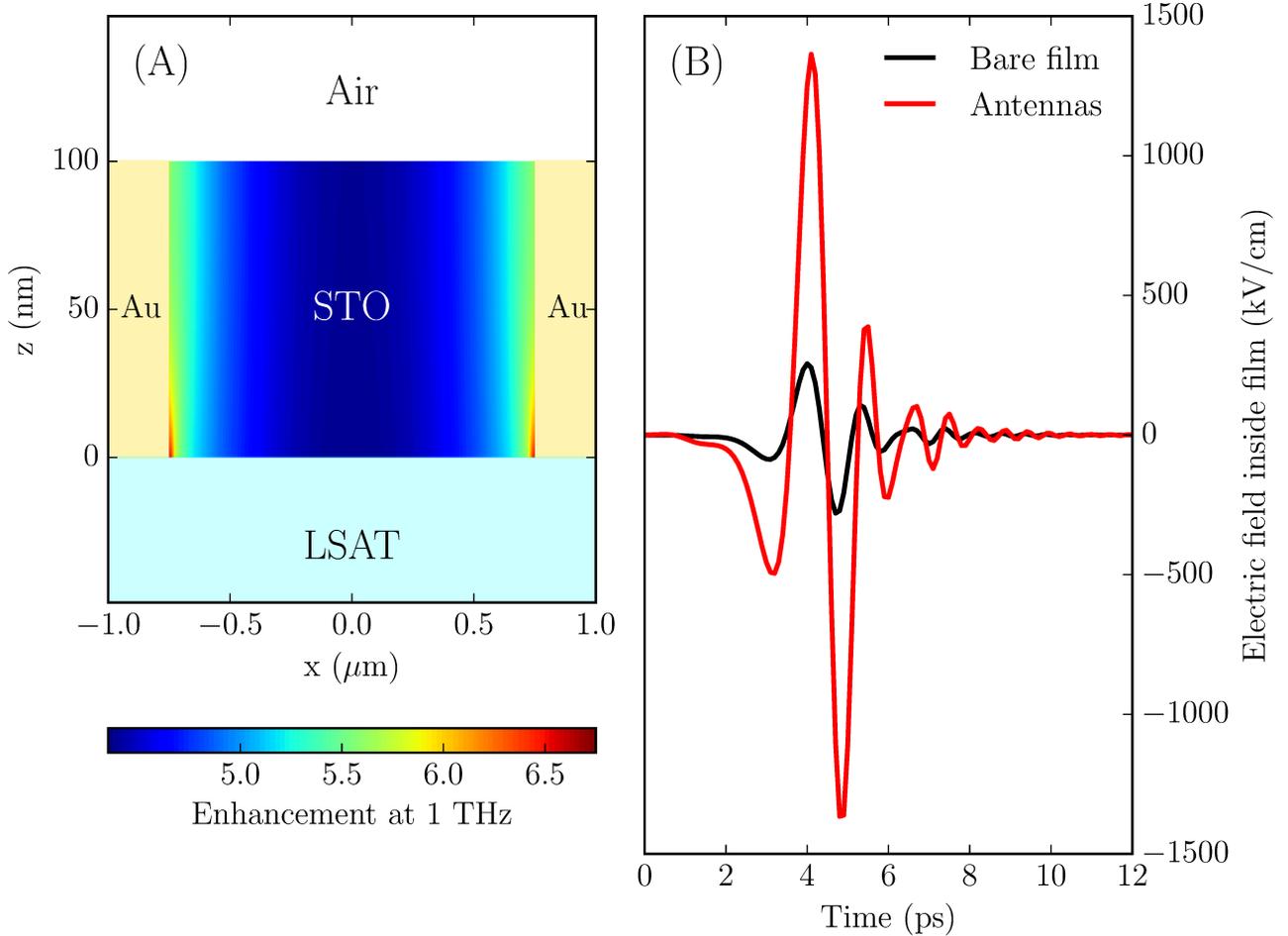

Fig. 2. (A) Cross-section view of the sample geometry overlaid with the THz enhancement inside the STO for a fixed frequency (1 THz). (B) Calculated waveform of THz radiation in the STO with (red) and without (black) metamaterial structures incorporating STO dispersion and measured free-space EOS.

X-ray pulses were generated at the Linac Coherent Light Source and the experiment was performed at the XPP end station[30] in monochromatic mode. The x-ray photon energy was 8 keV with a ~1 eV bandwidth after the monochromator. The x-ray pulses were 30 fs in duration at a repetition rate of 120 Hz. The diffraction measurements were collected with an x-ray pixel array detector (CS-PAD 140k detector[31]). For all measurements we oriented the sample to the top of the respective Bragg peak and integrated over a 2D slice of reciprocal space corresponding to the scattering peak. For every diffraction geometry, the THz radiation was incident collinearly with the x rays and was p-polarized with respect to the STO film. The x-ray spot size was 200 μm covering about 20 gaps of the structured film. Because of diffraction constraints, the incident angle of the x rays (and so the THz) for each crystal reflection was distinct. Specifically, the incidence angle referenced to the sample normal for the $(2\bar{2}3)$, $(\bar{2}23)$, and (004) reflections were respectively: 8°, 79°, and 38°. However, because of the large dielectric constant of LSAT[27] in the THz regime, we expected that the THz electric field would lie primarily in the plane of the STO film along $[1\bar{1}0]$ independent of incident angle because of refraction. Temperature regulation at 100 K was provided by a nitrogen cryostream (Oxford Instruments Cryojet 5). The sample temperature thus has a lower bound of 100 K but could be at most 10 K higher.

In Fig. 3A, we present the transient change in diffracted intensity for a bare STO film (no metamaterial structure) for three crystal reflections: $(2\bar{2}3)$, $(\bar{2}23)$, and (004). The THz free-space EO sampling data is overlaid for comparison (black). We subtract the average scattering signal before the THz arrives ($I_{t<0}$) from the intensity at time $t$ ($I_t$) and normalize by $I_{t<0}$. The fractional intensity change shows antisymmetric behavior for the $(2\bar{2}3)$ and $(\bar{2}23)$ peaks, while there is little time-resolved change in the (004) peak intensity. The ions of the STO film follow the THz electric field and modulate the structure factor, hence all atomic motion will be along the direction of the THz field polarization, namely in-plane along $[1\bar{1}0]$, with no atomic motion along the cross-plane [001] direction. As the THz radiation wavelength is roughly a factor of $\sim 10^6$ larger than the STO unit cell, adjacent cells will see essentially the same field and so we expect no short-time changes in unit cell volume. In other words we expect to couple only to volume-preserving, zone-center optical phonon modes in the STO. Note that the (004) structure factor is sensitive only to deviations along the cross-plane direction (parallel to [001]), and thus no transient signal from structure factor modulations for this peak is expected. However, there is significant in-plane momentum transfer for both the $(2\bar{2}3)$ and $(\bar{2}23)$ peaks, and thus we expect to see transient changes in the scattered intensity resulting from a modulation of the structure factor due to ionic motion. Moreover, the in-plane momentum transfer of the $(2\bar{2}3)$ and $(\bar{2}23)$ reflections are opposite in sign. Therefore, depending on the sign of the THz polarization, the electric field is parallel to either the $(2\bar{2}3)$ or the $(\bar{2}23)$ in-plane momentum transfer, and antiparallel to the other. For small ionic motions, this difference results in an increase in scattering signal for one peak and a decrease in the other, precisely what we measure in Fig. 3A. This result confirms that the transient signal in the diffracted intensity is dependent on the *field* of the driving THz, not its intensity, because only a field-driven process would yield opposite behavior for the two reflections.

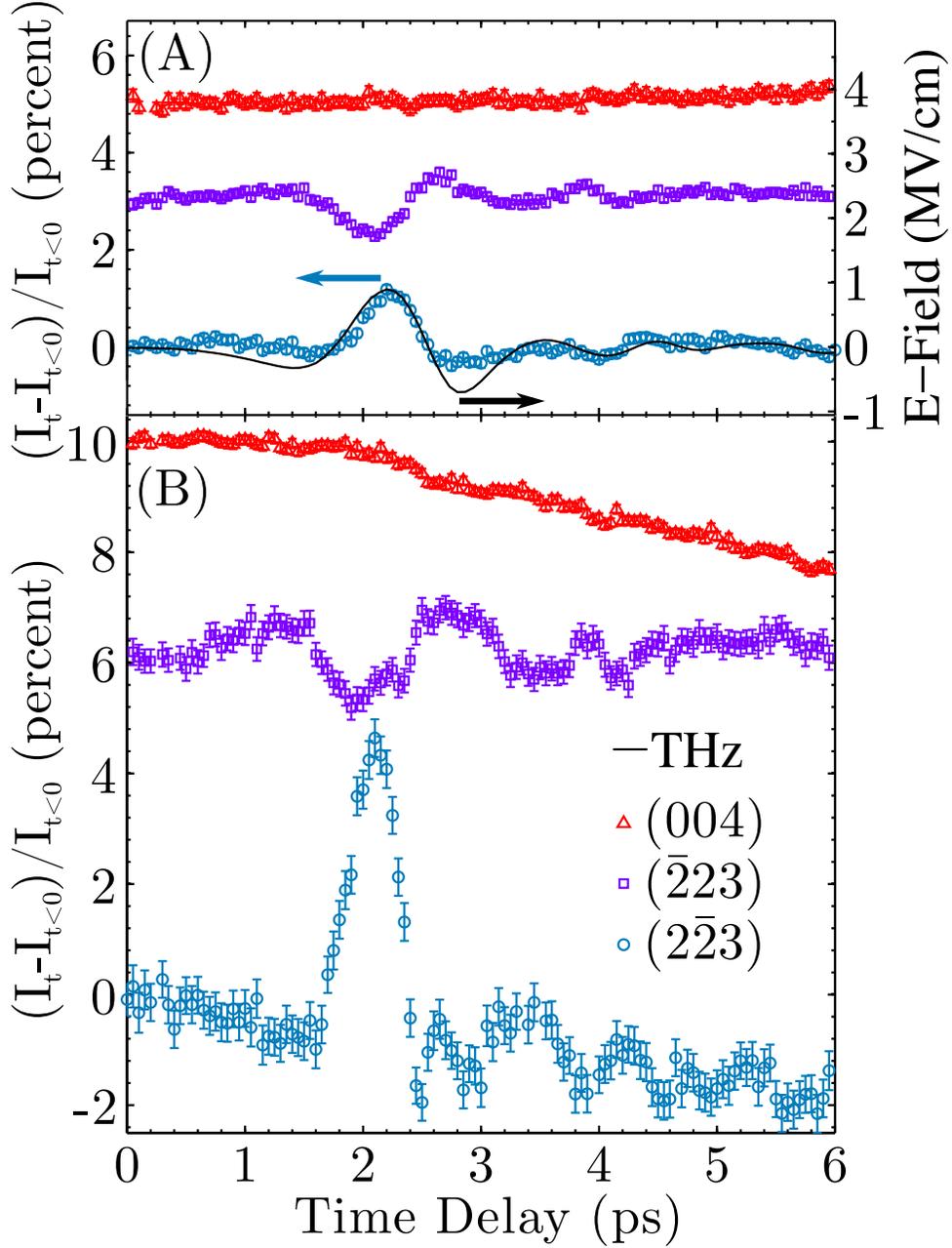

FIG. 3. Time-resolved Bragg peak intensity change for the (004), (2$\bar{2}$3), and ($\bar{2}$23) diffraction peaks (red triangles, purple squares, blue dots, respectively) at 100 K and 900 kV/cm for the bare sample (A) and the sample with metamaterial (B). Note all error bars in (A) and those on the red triangles in (B) are smaller than the symbol size. Traces are offset from zero for clarity. The EOS THz trace in free space is overlaid in (A).

In Fig. 3B, we show similar measurements for the sample with the metamaterial structure. Here we see that while the ($\bar{2}$23) reflection shows a peak change in scattering intensity of approximately 1%, comparable to the signal from the bare sample in Fig. 3A, the (2$\bar{2}$3) reflection reveals a five times larger change. The THz pulse was at a much more grazing angle (11° from the surface) for the ($\bar{2}$23) reflection compared to the near-normal (82° from surface) geometry for the (2$\bar{2}$3) peak. Due to the relatively large dielectric constant of the substrate this had little effect in the bare film. In contrast, the effect on the field enhancement is drastic because the effective microstructure spacing is modified. At near grazing, the microstructure array is stretched in space, and any field enhancement is strongly diminished. However, for the near-normal geometry, the microstructure can function as designed. Hence we expect to see a large enhancement for the ($2\bar{2}3$) reflection but relatively little change compared to the bare sample for the ($\bar{2}23$) reflection. This is what we indeed observe in Fig. 3B. We also conclude that the process of applying the microstructures did not substantially alter the STO phonon coupling to the THz via a change in strain because the signal in the ($\bar{2}23$) reflection remained unchanged, and the coupling is independent of x-ray reflection.

In addition, we show the (004) reflection with the metamaterial present in Fig. 3B. Here we see little change at early times but a slow decrease in signal on the few picosecond timescale, after the THz pulse has largely departed, compared to no clear change for same peak without metamaterial (Fig. 3A). We attribute this to heating in the film caused by absorption of field-enhanced THz fields leading to a decrease in peak intensity via the Debye-Waller factor[32] as well as possibly strain shifting the Bragg condition. This slow decrease is also seen in the background of the field-enhanced ($2\bar{2}3$) reflection, which has a similar magnitude momentum transfer (hence similar Debye-Waller factor) to (004) and a significant cross-plane component sensitive to longitudinal strain.

Note that the error bars for the metamaterial sample are in general larger than for the bare sample because the diffraction peak intensities are reduced. For example, the bare ($2\bar{2}3$) static reflection intensity was roughly twenty times stronger than the same reflection for the metamaterial sample. This is a larger difference than simple geometric arguments would suggest (we removed 85% of the sample to make the metamaterial covering), suggestive that there may be additional factors diminishing the metamaterial scattering intensity such as shadowing. However, the decrease in scattering strength did not substantially inhibit our ability to measure the enhancement of the transient signal induced by enhanced THz fields.

To further explore the effect of the microstructure field enhancement, we attenuated the incident THz field and measured the diffraction signal of the ($2\bar{2}3$) peak as a function of incident THz power with the metamaterial in place. In Fig. 4, we present the transient change in diffraction intensity for three THz field strengths. The values in the legend are estimated from free-space electro-optic sampling taken at the sample position and do not take into account the field enhancement effect. There is a monotonic increase in the maximum change in scattering intensity as a function of applied THz field. We show this change as a function of THz field strength, along with a linear fit constrained to pass through the origin, in the inset to Fig. 4. This fit follows the data within the error bars but does not exclude the possibility of a nonlinear response. There are two factors at work that may affect the linearity of the peak change in scattering. First the change in structure factor (and so scattered intensity) for small ionic motion is linear but has a non-negligible quadratic contribution for larger structural deviations. Second, the ionic potential of STO is known to have nonzero quartic contribution[33,34] and so for large enough driving fields the ionic motion will no longer be linear in the field.

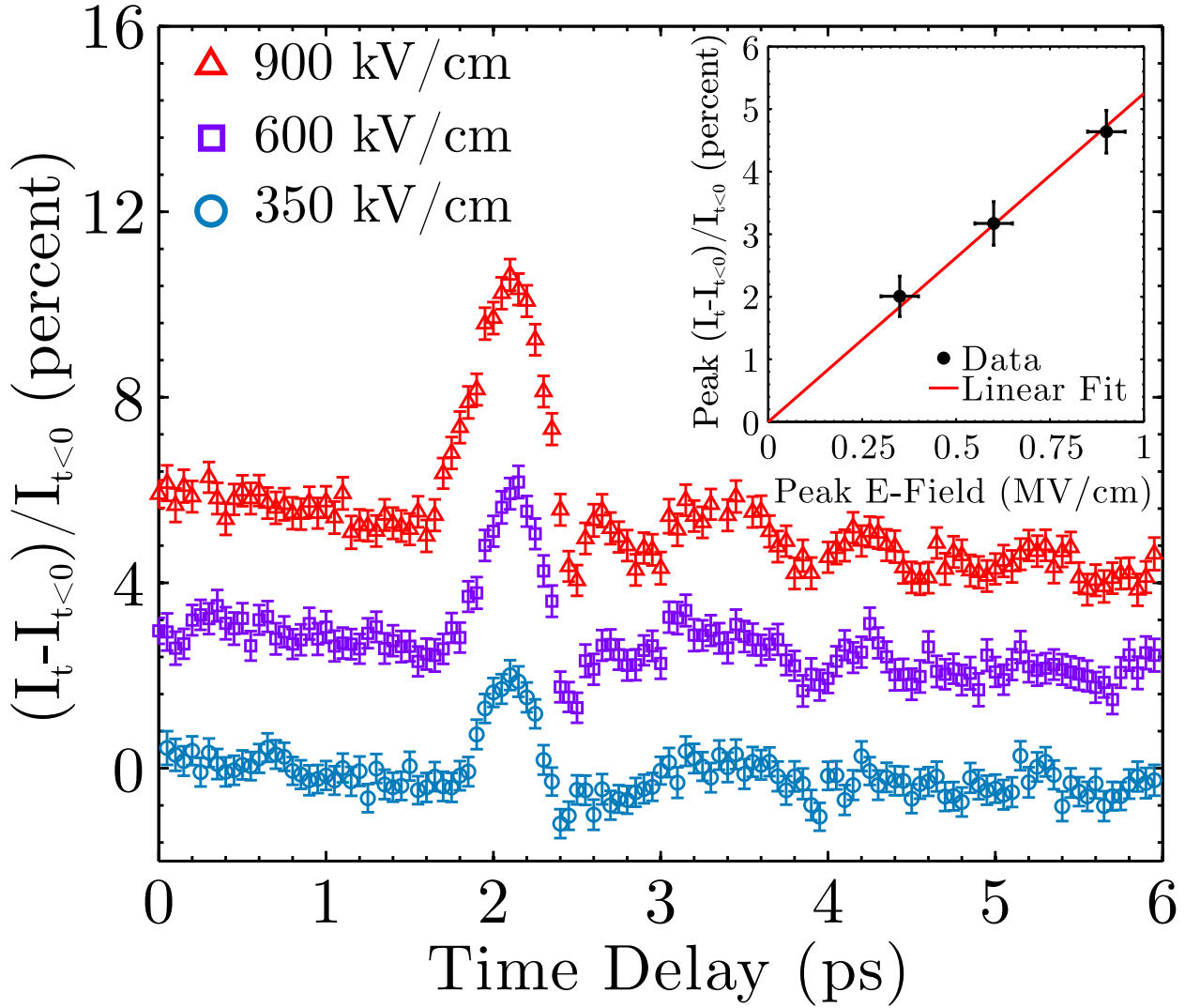

FIG. 4. Time-resolved change in Bragg peak intensity for the (2$\bar{2}$3) peak at 100 K as a function of incident peak THz electric field for the microstructured sample. Traces are offset from zero for clarity. The inset shows the peak intensity change for each applied THz field with a corresponding linear fit constrained to pass through the origin.

In summary, we have shown the viability of local THz field enhancement utilizing gold microstructures for time-domain x-ray diffraction experiments. Through measurements of several diffraction peaks we were able to confirm the field-driven nature of the THz excitation. Moreover we measured up to a factor of five increase in transient scattering signal (related to peak THz field) in the presence of the microstructures compared to bare samples. Last we confirmed that our system remained in a linear regime, suggestive that the THz waveform did not undergo extensive temporal shaping from the microstructure beyond amplitude scaling and phase shift (see Fig. 2B). This method is widely applicable to THz pump/x-ray probe experiments at x-ray free electron lasers and synchrotron sources and therefore should find broad use in structural studies of THz dynamics.

This work is supported by the Department of Energy, Office of Science, Basic Energy Sciences, Materials Sciences and Engineering Division, under Contract No. DE-AC02- 76SF00515 and by the NCCR Molecular Ultrafast Science and Technology (NCCR MUST), a research instrument of the Swiss National Science Foundation (SNSF). M.K. and M. C. H. are supported by the U.S. Department of Energy, Office of Science, Office of Basic Energy Sciences under Award No. 2015-SLAC-100238-Funding. The use of the LMN facilities at PSI are duly acknowledged as well as the support of V. Guzenko and A. Weber. M.P. acknowledges support from the Basque Government (Project n. PI2015-1-19) and from MINECO (Project n. FIS2015-64519-R, Grant n. BES-2013-063690 and n. EEBB-I-16-10873).